\documentclass[twoside,twocolumn,english,aps,showpacs,prl,superscriptaddress]{revtex4-1}
\usepackage[T1]{fontenc}
\usepackage[latin9]{inputenc}
\usepackage{color}
\usepackage{bm}
\usepackage{amsmath}
\usepackage{amssymb}
\usepackage{graphicx}
\usepackage{esint}
\usepackage{mathrsfs}

\makeatletter

\makeatletter
\newcommand*{\rom}[1]{\expandafter\@slowromancap\romannumeral #1@}
\makeatother

\usepackage{times}{\Large }

\makeatother

\usepackage{babel}
\begin{document}

\title{Topological Kondo Device for distinguishing Quasi-Majorana and Majorana signatures}

\author{Donghao Liu}

\affiliation{State Key Laboratory of Low Dimensional Quantum Physics, Department of Physics, Tsinghua University, Beijing, 100084, China}

\author{Zhan Cao}
\affiliation{Beijing Academy of Quantum Information Sciences, Beijing 100193, China}

\author{Xin Liu}
\affiliation{School of Physics and Wuhan National High Magnetic Field Center, Huazhong University of Science and Technology, Wuhan, Hubei 430074, China}

\author{Hao Zhang}
\affiliation{State Key Laboratory of Low Dimensional Quantum Physics, Department of Physics, Tsinghua University, Beijing, 100084, China}
\affiliation{Beijing Academy of Quantum Information Sciences, Beijing 100193, China}
\affiliation{Frontier Science Center for Quantum Information, Beijing 100184, China}

\author{Dong E. Liu}
\email{Corresponding to: dongeliu@mail.tsinghua.edu.cn}
\affiliation{State Key Laboratory of Low Dimensional Quantum Physics, Department of Physics, Tsinghua University, Beijing, 100084, China}
\affiliation{Beijing Academy of Quantum Information Sciences, Beijing 100193, China}
\affiliation{Frontier Science Center for Quantum Information, Beijing 100184, China}

\begin{abstract}
To confirm the Majorana signatures, significant effort has been devoted to distinguishing between Majorana zero modes (MZMs) and spatially separated quasi-Majorana modes (QMMs). Because both MZMs and QMMs cause a quantized zero-bias peak in the conductance measurement, their verification task is thought to be very difficult. Here, we proposed a simple device with a single nanowire, where the device could develop clear evidence of the topological Kondo effect in the topologically trivial phase with four QMMs. On the other hand, in the topological superconducting phase with MZMs, the transport signatures are significantly different. Therefore, our scheme provides a simple way to distinguish Majorana and quasi-Majorana modes.
\end{abstract}

\pacs{}

\date{\today}

\maketitle


{\em Introduction.--}
The topological superconductors can host localized zero-energy excitations named as \textquotedblleft Majorana
zero modes\textquotedblright{} (MZMs)~\cite{ReadGreen,1DwiresKitaev}. Among many experimental searches for MZMs, the semiconductor nanowire in proximity to an s-wave superconductor~\cite{Sau10,LutchynPRL10,1DwiresOreg,Sau10,Alicea10,CookPRB'11,AliceaRev,Mourik2012,Rokhinson2012,Deng2012,Das2012,Churchill2013,Finck2013,Nadj-Perge14,Albrecht16,deng2016Majorana,Zhang2017Ballistic,ZhangNN2018}
proved to be one of the most promising platforms to study non-Abelian braiding statistics~\cite{NonAbelian77,NonAbelian89,Ivanov2001NonAbelian,TQCreview} and topological quantum computation~\cite{kitaev,TQCreview}. Usually, a quantized zero-bias peak in the tunneling spectroscopy was considered as a smoking gun signature for the MZMs ~\cite{Law09}. However, many recent works have shown that individual near-zero-energy Andreev bound states (ABSs) can also cause a zero-bias conductance anomaly~\cite{kells2012PRB,TewarQM2013,Cayao-QDABS2015,Klinovaja2015,SanJose-ABS2016,CXLiuABSMZM,Setiawan-ABS17,Fernando-QDABS2018,reeg2018zero,moore2018twoTerminal,moore2018quantized,quasiMajoranaWimmer2019,pan2020physical,pan2020generic}.
If the potential near the nanowire's edge is smooth~\cite{kells2012PRB}, this ABS decomposes into two almost decoupled MZMs~\cite{moore2018twoTerminal,moore2018quantized,quasiMajoranaWimmer2019}. Because such states are in the topologically trivial phase, the two decomposed MZMs are also called ``quasi-Majorana Modes'' (QMMs)~\cite{quasiMajoranaWimmer2019}. In the tunneling spectroscopy experiments, only one of the two QMMs couples to the outside metallic lead, resulting in also a robust quantized zero-bias conductance peak~\cite{moore2018quantized,pan2020generic}. Therefore, it is tough to distinguish between QMMs and real MZMs in the local quantum transport experiments~\cite{moore2018quantized,moore2018twoTerminal,Donghao2020}.

With the rapid progress of the Majorana search in the past years, significant effort has been devoted to demonstrating non-Fermi liquid
(NFL) correlations due to the topologically protected Majorana
degrees of freedom~\cite{BeriPRL2012,Altland'13,BeriPRL2013,crampe2013kondo,tsvelik2013majoranakondo,galpin2014conductance,zazunov2014transport,
tsvelik2014topological,altland2014bethe,altland2014multichannel,eriksson2014non,eriksson2014tunnelingKondo,kashuba2015topologicalkondo,buccheri2015thermodynamics,
plugge2016kondo,herviou2016many,buccheri2016holographic,giuliano2016four,giuliano2016junction,zazunov20176pi,beri2017exact,bao2017topological,michaeli2017electron,
landau2017two,latief2018screening,snizhko2018parafermionic,gau2018Majoranabox,papaj2019multichannel,Roman2020signatures}.
The seminal work in Ref.~\cite{BeriPRL2012} have proposed an elegant idea of realizing a robust NFL Kondo effect~\cite{Kondo'64}, or the topological
Kondo effect (TKE), by using the topological degeneracy that arises from the non-local MZMs. Such TKE in its minimal setup~\cite{BeriPRL2012,Altland'13,BeriPRL2013} consists of a floating topological superconducting island supporting four localized MZMs
($M_{tot}=4$), three of which ($M=3$) are tunnel coupled to the single-channel conducting leads. The critical experimental phenomenon is
that as the temperature decreases, the linear conductance will saturate at a fractional value ($G=2e^{2}/Mh$)~\cite{BeriPRL2012}, showing a crossover from a weak coupling trivial fixed point to a strong coupling NFL fixed point. Besides, we believe that the measurement of TKE could both provide strong evidence of coherence nature in Majorana devices (alternative methods like Majorana teleportation interferometer~\cite{MajoranaTeleportation} and dissipative Majorana teleportation~\cite{Donghao2020}) and a transport characterization scheme for Majorana qubits.

\begin{figure}
\includegraphics[width=0.8\columnwidth]{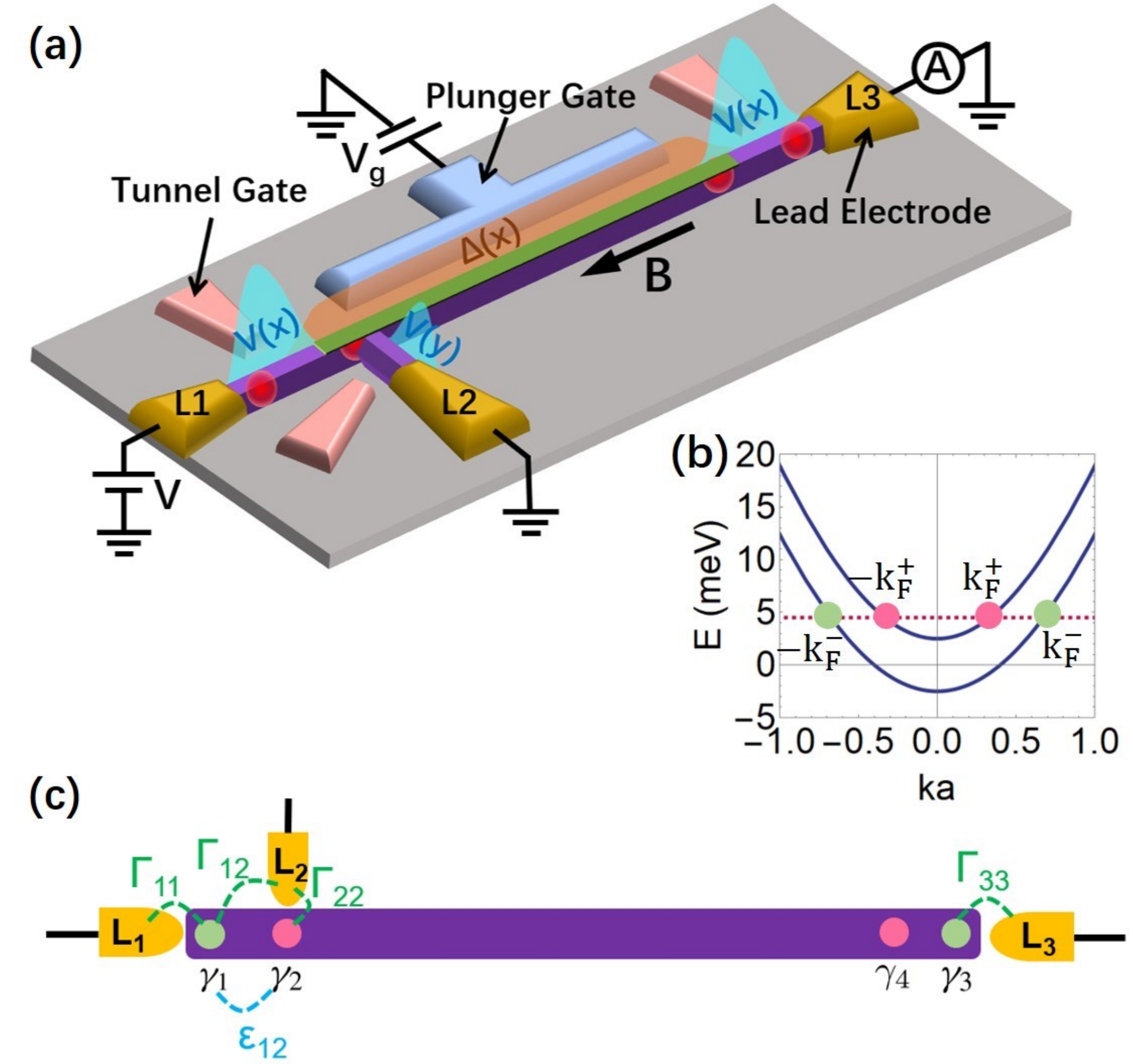}\caption{\label{fig:setup}(a) is the proposed experiment setup to study
the TKE in quasi-Majorana nanowire. 
(b) is the energy dispersion of a quasi-one-dimensional nanowire without the superconductor shell.
(c) is a schematic of this three-leads setup. The most influential hybridizations
are labeled. }
\end{figure}

A single Majorana nanowire system with smooth potentials at both ends could host four QMMs, which satisfy the minimum requirements for TKE. Therefore, we would like to ask if the TKE can be realized and observed experimentally in such a single-nanowire setup.
In this work, we propose a simple quasi-Majorana device with a single nanowire as shown in Fig. \ref{fig:setup} to realize TKE in the topologically trivial phase, and derive the conditions under which TKE could appear and potentially be observed in experiments. We believe that our device is much easier to realize for experimentalists than the standard TKE devices that require at least two nanowires. Besides, our device provides a robust experimental scheme to distinguish between topological MZMs and non-topological QMMs.


{\em Proposed device structure.--} 
The proposed system setup is shown in Fig. \ref{fig:setup}(a). A
semiconductor nanowire (purple) with Rashba spin-orbit coupling is
in proximity to an $s$-wave superconductor shell (green). A magnetic
field $B$ is applied in parallel with the nanowire. We consider a floating island with an Coulomb electrostatic energy
$U_C=E_{C}\left(n-n_{g}\right)^{2}$, where $n$ is the total electron number in the island. The plunger gate control the parameter $n_{g}$ with $n_{g}=CV_{g}/e$, where $C$ is the effective capacitance and $V_{g}$ is the gate voltage. The
tunnel-gates control the couplings between the leads and the nanowire
and give rise to the smooth potentials [$V(x,y)$]. The proximity-induced gap $\Delta\left(x\right)$ gradually
vanishes in the regime where no superconducting shell is covered. There is a T-shape structure near the left side of the nanowire, and this  T-junction can be realized from the epitaxial growth~\cite{Nature2017Network} or the selective area growth~\cite{krizek2018field,vaitiekenas2018selective,het2020plane}. As shown in Fig. \ref{fig:setup}(c), three lead electrodes ($L1$-$L3$)  cover the wire ends to detect the QMMs ($\gamma_1$-$\gamma_3$).  
The $L2$ lead covers almost the whole side-leg to strengthen the $\gamma_{2}-L2$ coupling and smoothen the potential near the connection point. The $L2$ lead is also magnetized using a ferromagnetic lead or by a normal metallic lead in proximity to a magnetic
insulator. Each pair of QMMs at the same wire-end always has the opposite spin polarization~\cite{quasiMajoranaWimmer2019,stanescu2019robust}. Therefore, the magnetization of $L2$ further suppresses the potential "crosstalk" between the  $\gamma_{1}$ and the $L2$ lead.

\begin{figure}
\includegraphics[width=1\columnwidth]{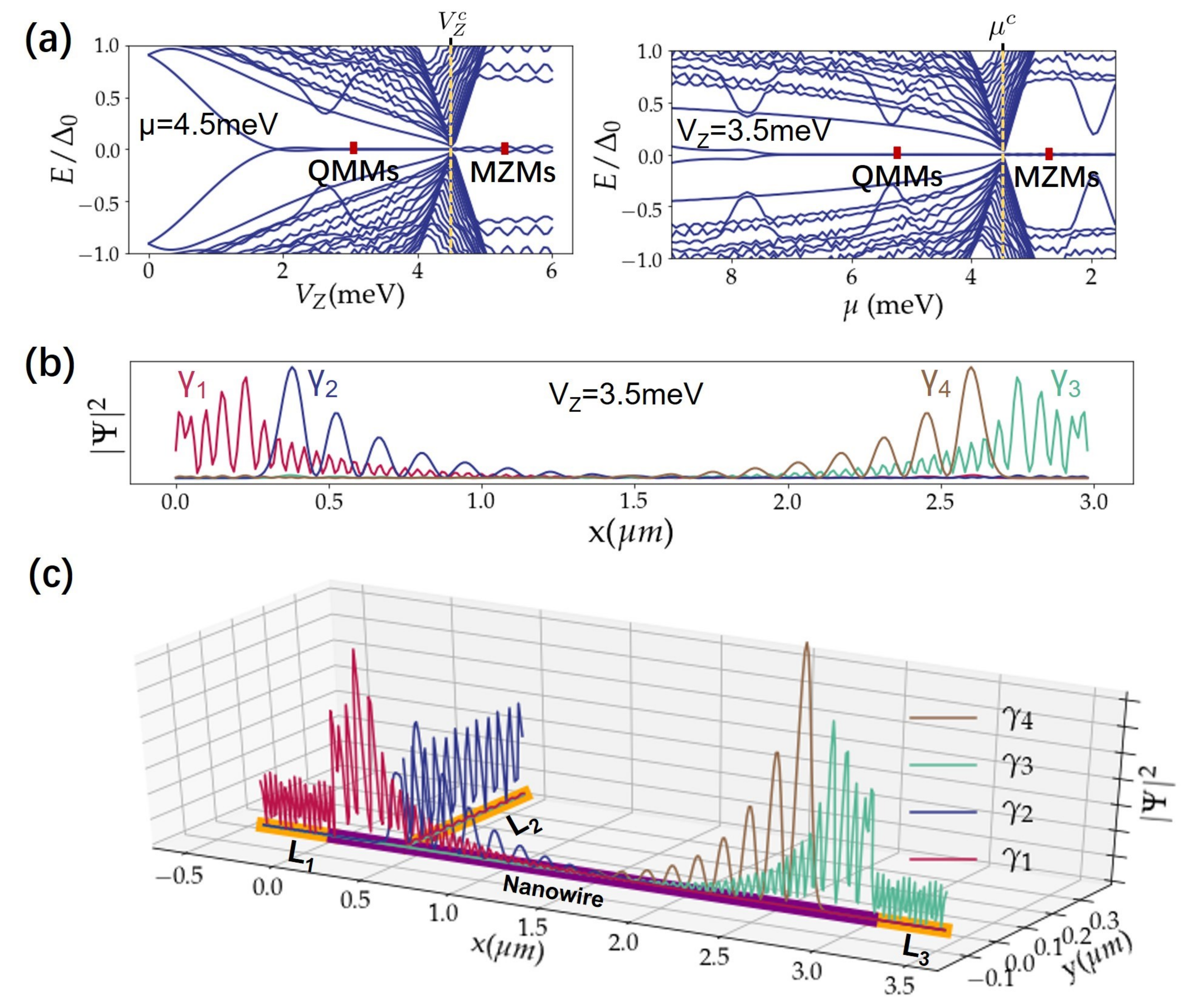}
\caption{\label{fig:QSMWF} (a) The energy spectrum of a bare quasi-Majorana nanowire with smooth potential on one side as a function of the Zeeman energy $V_Z$ and the chemical potential $\mu$. The critical $V_Z^{c}$ and $\mu^c$ at which topological phase transition happens are labeled. (b) The wave functions of four QMMs along the nanowire. (c) The wave functions of the four QMMs when all three leads are attached. In (a)-(c), the chemical potential of the nanowire $\mu$ is tuned to $4.5\text{meV}$ and in (b), (c) $V_{Z}$ is tuned to $3.5\text{meV}$. The other parameters of the nanowire are taken as:  $m^{*}=0.023m_{e}$, $\alpha=50\text{meVnm}$, the induced gap $\Delta_{0}=0.5\text{meV}$ and the lattice constant $a=10\text{nm}$. The parameters for the variation of the potential and the superconducting gap are $V_{0}=6\text{meV}$, $x_{V}=200\text{nm}$, $\sigma_{V}=100\text{nm}$, $x_{\Delta}=250\text{nm}$ and $\sigma_{\Delta}=100\text{nm}$.}
\end{figure}

{\em TKE in quasi-Majorana wire.--} 
It is known that four QMMs could be generated in a single
nanowire with smooth potentials~\cite{kells2012PRB,moore2018twoTerminal,moore2018quantized,quasiMajoranaWimmer2019} at both sides. 
The one-dimensional Bogoliubov-de Gennes (BdG) Hamiltonian of a Majorana nanowire extending
in $x$ direction can be written as
\begin{equation}
H=\left(\frac{p_{x}^{2}}{2m^{*}}-\mu+V(x)\right)\tau_{z}-\alpha p_{x}\sigma_{y}\tau_{z}+V_{\mathrm{Z}}\sigma_{x}+\Delta(x)\tau_{x},\label{eq:BDGH}
\end{equation}
where $p_{x}$ is the momentum, $m$$^{*}$ is the effective mass,
$\mu$ is the chemical potential, $V$ is the electrostatic potential,
$\alpha$ is the spin-orbit coupling (SOC) strength, $V_{\text{Z}}$
is the Zeeman energy due to the magnetic field parallel to the nanowire,
and $\Delta$ is the proximity-induced superconducting gap. $\sigma_{i}$ and $\tau_{i}$ $\left(i=x,y,z\right)$
are Pauli matrices which act on spin and particle-hole space, respectively.
Here we use a Gaussian shape $V\left(x\right)=V_{0}\exp\left[\left(x-x_{V}\right)^{2}/\sigma_{V}^{2}\right]$ to model the smooth
potential for both left and right junctions {[}see Fig.~\ref{fig:setup}(a){]}. The transition between the superconducting and non-superconducting regimes is also smooth: 
$\Delta\left(x\right)=\Delta_{0}\left\{ 1+\tanh\left[\left(x-x_{\Delta}\right)/\sigma_{\Delta}\right]\right\} $.
The energy spectrum as a function of the Zeeman energy and chemical potential of this system
is shown in Fig. \ref{fig:QSMWF}(a). MZMs appear in the topological regime.
In the topologically trivial regime $(V_{Z}<\sqrt{\Delta_{0}^{2}+\mu^{2}})$
there are two Andreev bound states stick to zero energy which can be decomposed
into four QMMs, and their wave functions are shown in  Fig.~\ref{fig:QSMWF}(b).

The formation of quasi-Majoranas requires that the Fermi surface intersects two helical bands of the nanowire, forming two pairs of Fermi points as shown in Fig.\ref{fig:setup}(b). Although the system is in the topologically trivial regime, if the scatterings between the two pairs of Fermi points $\pm k_F^+$ and $\pm k_F^-$ are weak enough (e.g. due to the smooth potential), the two bands can be seen as two independent spinless bands~\cite{kells2012PRB}. In the presence of the proximity-induced superconducting correlation, the proximity effect induces Cooper correlations for electrons near each pair of Fermi points and generates two Bose condensates.
Each condensate bears two QMMs, for which the spatial wavefunction is shown in Fig.~\ref{fig:QSMWF}(b). The inner two QMMs belong to one condensate and the outer two QMMs belong to the other. Then with $M=3$ leads coupled to three QMMs, the tunneling Hamiltonian can be written as
\begin{equation}
H_{T}=\sum_{j}t_{jj}\gamma_{j}\psi_{j}e^{i\phi_{j}/2}+H.c.,
\end{equation}
where $t_{jj}$ is the lead-QMM tunneling amplitude, $\psi_{j}$ is the electron annihilation operator of lead-$j$ ($j=1,2,3$).
$\phi_{1,3}=\phi_A$ and $\phi_{2,4}=\phi_B$ represent the phase of the outer ($A$) and inner ($B$) condensates, which are conjugate to the number of Cooper pairs of each condensates $n_A$ and $n_B$: $\left[\phi_\alpha, \hat{n}_\alpha\right]=i\text{ },\alpha=A,B$. If $t_{jj}\ll E_C$ and $n_g$ is tuned to the Coulomb blockade valley, using the Schrieffer-Wolff transformation~\cite{hewson1997kondo,BeriPRL2012}, one can obtain the effective exchange Hamiltonian~\cite{kells2012PRB}:
\begin{equation}
H^T_{\text{eff}}=\sum_{j\neq k}\frac{t_{jj}t_{kk}^{*}}{E_{C}}\gamma_{j}\gamma_{k}\psi_{k}^{\dagger}\psi_{j}e^{i\left(\phi_{j}-\phi_{k}\right)/2}.
\label{eq:HTeff}
\end{equation}
 If $\gamma_j$ and $\gamma_k$ belong to different condensates there is phase exponential factor $e^{\pm i\left(\phi_{A}-\phi_{B}\right)/2}$ which represents the electron transfer between the two condensates. For our quasi-Majorana device, both condensates exchange Cooper pairs with the proximity-superconductor at the same spatial positions; and therefore, the effective coherent Josephson coupling $H_J=-E_J \cos\left(\phi_A-\phi_B\right)$ between the two condensates could be very strong. For a large $E_J$, $\phi_A-\phi_B$ will be fixed at $0$. In addition, the two condensates together share the same spatial locations and only feel a single constant charging energy. Then $H^T_{\text{eff}}$ is reduced to the ideal topological Kondo model~\cite{BeriPRL2012}. Here, we note that, in order to observe the TKE in the Majorana double wire `H-shape'' qubits~\cite{ScalableDesign,ZhangLiuReview} (the simplest device to realize four MZMs using topological wires), we need 1) the Cooper pairs in different topological wires to feel the same charging energy, 2) the Josephson coupling between the two different wires is very strong. Therefore, our proposal provides a more natural platform for TKE.

\begin{figure}
\includegraphics[width=1\columnwidth]{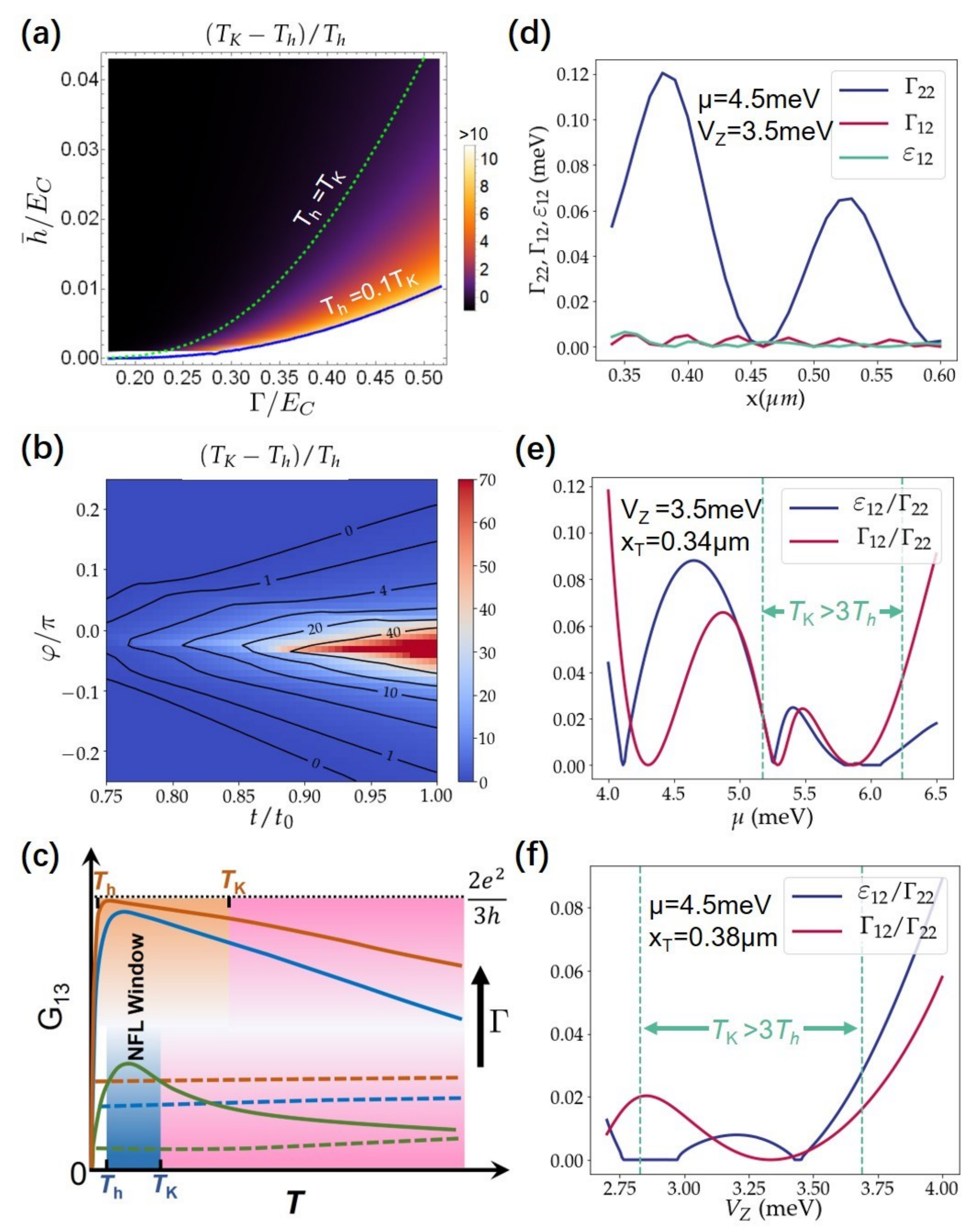}\caption{\label{fig:T_results}
(a) The relative size of the NFL window $\left(T_\mathrm{K}-T_{\mathrm{h}}\right)/T_{\mathrm{h}}$ as a function of the local Majorana-lead hybridization and the effective Majorana-Majorana hybridization. 
(b) The simulation result which gives  the dependence of $\left(T_\mathrm{K}-T_{\mathrm{h}}\right)/T_{\mathrm{h}}$ on the tunnel-coupling strength and the magnetization direction of lead-$2$. Relevant parameters used are $x_{\mathrm{T}}=0.38\mu \text{m}$, $\mu=4.5\text{meV}$, $V_Z=3.5\text{meV}$, $E_C=0.2\text{meV}$ and $\nu=10\text{meV}^{-1}$.
(c) The conductance $G_{13}$ for the setup shown in Fig.~\ref{fig:setup} as a function of temperature. The solid (dashed) lines describes the topological trivial case with QMMs (topological case with MZMs). The curves with the same local Majorana-lead hybridization $\Gamma$ are drawn in the same color. The upper curves correspond to larger $\Gamma$. The blue (orange) shaded area indicates the NFL (Kondo enhancement) window.
(d) The couplings $\Gamma_{22}$, $\Gamma_{12}$ and $\varepsilon_{12}$ when $L2$ lead is attached at different position $x$ measured from the left side of the nanowire.
(e) and (f) show ratios of the hybridization parameters $\Gamma_{12}/\Gamma_{22}$ and  $\varepsilon_{12}/\Gamma_{22}$ as a function of $\mu$ and $V_Z$.
The range of $\mu$ and $V_Z$ are chosen to support QMMs as shown in Fig.~\ref{fig:QSMWF}(a). 
}
\end{figure}

{\em Conditions for TKE.--} Here we will review the conditions for experimental observation of TKE 
in a floating Majorana island with $N$ MZMs coupling to $M$ ($\leq N$) different metallic leads. 
The Kondo temperature $T_{\mathrm{K}} \simeq E_{c} \exp [-\pi E_C/ 2(M-2) {\Gamma}]$ describes the crossover energy scale between the trivial Fermi Liquid (FL) fixed point and the NFL fixed point~\cite{hewson1997kondo,BeriPRL2012,Altland'13,BeriPRL2013} . 
 $\Gamma$ is the average value of the Majorana level broadenings $\Gamma_{jj}$ over all  $j$ leads, with $\Gamma_{jj}=2\pi\nu \left| t_{jj}\right|^{2}$ where $\nu$ is the density of states (DOS) at the fermi level of the leads. However, there are ``imperfect terms'' like a mutual hybridization between two MZMs: $i\varepsilon_{jk}\gamma_j\gamma_k$, or a nonzero crossed couplings between $\gamma_j$ and lead-$k$: $t_{j\neq k}$ as shown in Fig.~\ref{fig:setup}(c). If these imperfect terms are small ($\left|t_{jk}\right|,\varepsilon_{jk}\ll t_{jj}$, $j\neq k$), the TKE could exist but they give rise to another crossover energy scale $T_h$ ~\cite{galpin2014conductance,altland2014multichannel} below which these terms will drive the system away from the TKE regime of  $M$ MZMs to the TKE of $M-2$ MZMs. $T_h$ can be estimated~\cite{galpin2014conductance,altland2014multichannel} as $T_{\mathrm{h}}\simeq T_{\mathrm{K}}\left(\bar{h}/{T_{\mathrm{K}}}\right)^{M/2}$, where $\bar{h}$ is the typical value of the effective Majorana-Majorana hybridization: $\bar{h}=\max\left|h_{jk}\right|=\max\left|\varepsilon_{jk}+\sum_{p} \frac{t_{k p} t_{j p}^{*}}{i E_{\mathrm{C}}}\right|$~\cite{supp}, which include the effects from both two imperfect terms. For the case $M=3$, the system will be driven to a trivial FL state below $T_h$ . This will result in a transparent window of NFL state in the temperature range $T_{\mathrm{h}}\ll T\ll T_{\mathrm{K}}$.
The dependence of the NFL window size $\left(T_\mathrm{K}-T_\mathrm{h}\right)/T_\mathrm{h}$ on the Majorana-lead hybridization ${\Gamma}$ and the Majorana-Majorana hybridizations  $\bar{h}$  is shown in Fig.~\ref{fig:T_results}(a), which indicates that we have a large parameter regime to observe TKE ($T_\mathrm{h}\ll T_\mathrm{K}$).



{\em TKE in quasi-Majorana nanowires.--} 
In order to observe the TKE in our quasi-Majorana nanowire setup, we need to couple at least three QMMs to outside metallic leads as shown Fig. \ref{fig:setup}(c). We first check whether the device can satisfy the requirement for TKE.
To demonstrate how the lead couples to the QMMs, we study the wave function leakage into the leads and numerically calculate the spatial distribution of their wave functions throughout the device structure. 
Without attaching the leads, the wave functions of four QMMs are shown in  Fig. \ref{fig:QSMWF}(b), and we saw the partially separated QMMs $\gamma_1$ and  $\gamma_2$ ($\gamma_3$ and  $\gamma_4$) located closed to the left (right) end of the wire.  
After attaching leads to the nanowire, the wave function of  $\gamma_{1}$ and $\gamma_{3}$ show apparent leakage into the corresponding leads as shown in Fig. \ref{fig:QSMWF}(c), which could induce strong Majorana-lead couplings. While the leakage of  $\gamma_{2}$ ($\gamma_{4}$ ) into $L1$ lead ($L3$ lead) is significantly suppressed, and therefore, these crossed Majorana-lead couplings can be neglected. Considering a long wire and a very smooth barrier potential landscape, we can imagine that the direct Majorana-Majorana coupling is very small.  However, in the QMM nanowire, $\gamma_{1}$ and $\gamma_{2}$ are only partially separated in space, and there is some small but clearly visible contribution from $\gamma_{1}$ on top of the major $\gamma_{2}$ part as shown in Fig. \ref{fig:QSMWF}(b). Therefore, both $\gamma_{1}$  could leak into the attached $L2$ lead, and cause strong crossed Majorana-lead couplings. Because the $\gamma_{1}$ and $\gamma_{2}$ have the opposite spin polarization~\cite{quasiMajoranaWimmer2019,stanescu2019robust}, we consider a spin-polarized $L2$ lead to resolve this issue. The full spatial distribution of the wave function is shown in  Fig. \ref{fig:QSMWF}(c), which tells us all the "imperfect" couplings could be very small and thus have a chance to fulfill the TKE conditions.

Next, we quantitatively study Majorana-Majorana hybridization $\varepsilon_{ij}$ between $\gamma_i$ and  $\gamma_j$, and Majorana-lead hybridizations $\Gamma_{ij}=2\pi\nu \left|t_{ij}\right|^{2}$ between $\gamma_{i}$ and lead-$j$. Those hybridizations are labeled in in Fig. \ref{fig:setup}(c).  We have the situation that 1) the wave functions of $\gamma_{1}$ and $\gamma_{2}$ have a finite small overlap, and 2) the attached lead-$2$ connects the wire in their overlapping regime; and therefore, the most influential hybridization factors are $\Gamma_{12}$ and  $\varepsilon_{12}$. Other hybridizations can be safely neglected. According to the expressions of $T_{\mathrm{K}}$ and $T_{\mathrm{h}}$, we require small ratios $\Gamma_{12}/\Gamma_{22}$ and $\varepsilon_{12}/\Gamma_{22}$ to reach the conditions in general.
We numerically compute those hybridizations in the lattice model using a Kwant simulation~\cite{KwantWimmer,supp}.


The hybridization parameters are numerically shown in Figs. \ref{fig:T_results}(d)-(e). By changing the connection point $x_{\text{T}}$ between the lead-$2$ and the nanowire from left to right, the coupling strengths oscillate as shown in Fig. \ref{fig:T_results}(d). Those oscillations comes from the variation of the wave functions and the spin densities of QMMs~\cite{supp}. One can choose a range of $x_{\text{T}}$ near $\gamma_{2}$ such that $\Gamma_{22}\gg\Gamma_{12}$ and $\varepsilon_{12}$, where the lead-$2$ is only strongly coupled to $\gamma_2$. In a practical situation with a fixed $x_{\mathrm{T}}$, one can tune the value of $\Gamma_{22}$, $\Gamma_{12}$ and $\varepsilon_{12}$ by shifting the wave function horizontally, which can be achieved
by changing the chemical potential $\mu$ or the Zeeman energy $V_Z$. The ratios $\Gamma_{12}/\Gamma_{22}$ and $\varepsilon_{12}/\Gamma_{22}$ as a function of $\mu$ and $V_Z$ are plotted in Fig. \ref{fig:T_results}(e) and (f), which indicates a large regime to observe the TKE. Besides, the lead-Majorana hybridizations can also be tuned by the tunnel gate.
The dependence of the relative value $\left(T_\mathrm{K}-T_{\mathrm{h}}\right)/T_{\mathrm{h}}$ on the $L2$ lead-nanowire coupling $t$ and the $L2$'s magnetization direction angle $\varphi$  is shown in Fig.~\ref{fig:T_results}(b). Here, the magnetization direction is represented by the angle $\varphi$ through ($\cos \varphi\boldsymbol{\hat{x}}+\sin\varphi\boldsymbol{\hat{z}}$). The numerical result indicates a large parameter regime to support TKE.

{\em Quasi-Majorana vs Majorana.--} 
Finally, we discuss how to distinguish the Majorana modes from the quasi-Majorana modes
in our proposed setup shown in Fig.\ref{fig:setup}(a). In experiment, we can apply a voltage on $L1$ lead and detect the current in $Lk$ lead ($k=2,3$),
which yields the linear conductance $G_{1k}=\frac{dI_k}{dV_1}|_{V_1\rightarrow 0}$. $G_{1k}$ will show Coulomb blockade (CB) oscillations as we tune the gate voltage $V_{g}$. In order to observe the TKE, we fix $V_{g}$ to a certain value in the CB valley. 

If the nanowire is in the topological phase, there are only two MZMs ($\gamma_1$ and $\gamma_2$) located at each side of the nanowire. Then, MZM $\gamma_1$ couples to both the $L1$ and $L2$ leads, and  MZM $\gamma_2$ couples to the $L3$ lead. In the CB valley, the main contribution to the conductance comes from the electron cotunneling processes at low temperature and sequential tunneling processes at higher temperature~\cite{van2016PRB}. 
It is known that the conductance $G_{1k}$ is given by the Breit-Wigner formula~\cite{van2016PRB}: $G_{\mathrm{1k}}^{M}=\left(4T\right)^{-1}\int_{-\infty}^{\infty}d\xi g_{1k}\left(\xi\right)\cosh^{-2}\left(\xi/2T\right)$ with $g_{1k}\left(\xi\right)=\left(e^{2}/h\right)\Gamma_{1}\Gamma_{k}[\left(\xi-E_{C}\right)^{2}+(\sum_{l=1}^{3}\Gamma_{l})^{2}/4]^{-1}$, where $\Gamma_i$ denotes the hybridization strength between $Li$ lead and its nearby MZM. Because $G^{M}_{1k}$ depends on the value of $\Gamma_k$, $G^{M}_{12}$ and $G^{M}_{13}$ are not necessarily equal. We demonstrate the low-temperature $G^{M}_{13}$ as a function of $T$ for the case that the three leads are symmetrically coupled to the nanowire in Fig.~\ref{fig:T_results}(c) using dashed lines,  where the upper dashed line indicates the case with larger $\Gamma$. The cotunneling conductance is almost a constant for temperature much below the induced SC gap.

If the nanowire is in the topologically trivial phase with four QMMs, the system will show clear TKE as discussed before.  For the temperature regime $T_\mathrm{K}\ll T \ll E_C$, the conductance signature is induced by the cotunneling processes similar to the topological cases. For $T\sim T_\mathrm{K}$, the electron transports are significantly modified by the Kondo physics, and the conductance will show the Kondo enhancement with a logarithmic scaling behavior $G_{1k}\propto \ln^{-2}\left(T/T_{\mathrm{K}}\right)$~\cite{BeriPRL2012}. Further lowering the temperature below $T_{\mathrm{K}}$ into the NFL window, the $T$-dependence of the conductance will be characterized by the typical power-law behavior; and finally, the conductance will reach the fractional quantized value $2e^2/3h$ if the lead-QMM couplings are isotropic: $G_{13}/G_{12}\rightarrow 1$. However, when the temperature reaches the regime $T\sim T_h$, the system starts returning to the FL behavior. In this case, the coupling between $\gamma_1$ and $\gamma_2$  can't be neglected anymore; and therefore, the non-local conductance $G_{13}$  approaches zero at low $T$. Because $G_{12}$ is from the local transport via the fermionic state formed by $\gamma_1$ and $\gamma_2$, there is still a remaining value $G_{12}\sim \left(e^2/h\right)\left({\Gamma}_{11} {\Gamma}_{22} /E_C^2\right)$ at $T\rightarrow 0$ due to the elastic cotunneling. We illustrated those behaviors using the solid line in Fig.~\ref{fig:T_results}(c).

Therefore, the QMMs can be distinguished from the MZMs by the conductance measurement (the Kondo enhancement window or the NFL window with the fractional value) as we illustrated in  Fig.~\ref{fig:T_results}(c). It is also worth mentioning that by increasing the hybridization strength $\Gamma$ the logarithmic window will shrink, while the NFL window becomes larger. Considering two limits: 1) when $\Gamma$ is large close to $E_C$, the Kondo enhancement window will disappear but the NFL window will be still very large~\cite{Roman2020signatures}; 2) when $\Gamma$ is small, $T_{\mathrm{K}}\rightarrow T_{\mathrm{h}}$ and the conductance will turn down and decrease to $0$ before reaching the fractional value (e.g. a case shown in  the solid green line of Fig.~\ref{fig:T_results}(c)).

{\em Conclusions.--} 
In this work, we study the conditions for observing TKE in a quasi-Majorana
nanowire. When three quasi-Majoranas
strongly couple to three leads with the crossed couplings suppressed,
topological Kondo effect could appear. We proposed a simple experimental setup for observing topological Kondo effect in a single nanowire system, and our scheme could be applied to distinguish Majorana from quasi-Majorana systems.

\begin{acknowledgments}
The authors acknowledge the support from Tsinghua University Initiative Scientific Research Program, NSF-China (Grant No.11974198), and the startup grant from State Key Laboratory of Low-Dimensional Quantum Physics and Tsinghua University.
\end{acknowledgments}

\appendix

\begin{widetext}

\section*{Supplementary Material for\\``Topological Kondo Device for distinguishing Quasi-Majorana and Majorana signatures''}
In this supplementary material, we will provide some details about: ~\ref{section01}.) Details of the set-up of quasi-Majorana nanowire. ~\ref{section02}.) The tunneling conductance to the ferromagnetic lead.

\section{Details of the set-up of quasi-Majorana nanowire\label{section01}}

\subsection{Hamiltonian of the nanowire}
The Hamiltonian of the quasi-Majorana system $\int dx\psi^{\dagger}\left(x\right)H\left(x\right)\psi\left(x\right)$ with $H\left(x\right)$ shown in Eq.~(1) of the main text can be discretized on a one-dimensional atom chain:
\begin{equation}
\mathcal{H}_{NW}= \sum_{i}\left(\psi_{i}^{\dagger}\left\{[2 t-\mu+V(i)+V_Z \sigma_{x}] \tau_{z}+\Delta(i) \tau_{x}\right\} \psi_{i}\right.\left.-\left[\psi_{i+a}^{\dagger}\left(t+i \tilde{\alpha} \sigma_{y} \right)\tau_{z} \psi_{i}+\text { H.c. }\right]\right).\label{eq:dBDG_H}
\end{equation}
The Nambu basis has be changed from $\psi\left(x\right)=\left(c_{\uparrow x}, c_{\downarrow x}, c_{\uparrow x}^{\dagger}, c_{\downarrow x}^{\dagger}\right)$ to $\psi_{i}=\left(c_{\uparrow i}, c_{\downarrow i}, c_{\uparrow i}^{\dagger}, c_{\downarrow i}^{\dagger}\right)$ where $i$ labels the atom site and $i+a$ labels its nearest neighbour to the right. The hopping constant $t=\hbar^2/2 m^* a^2\approx13\text{meV}$ and the Rashba SOC strength $\tilde{\alpha}=\alpha/2a=2.5\text{meV}$. The valus of all the parameters can be found in the caption of Fig.~2 of the main text. The spatial distribution of the potential $V\left(i\right)$ and the induced gap $\Delta\left(i\right)$ is shown in Fig.~\ref{SI_QSMWF}(a). Solving the Eq.~(\ref{eq:dBDG_H}) of a $3 \mu m$ wire gives the spectrum and the wavefunctions of the QMMs and MZMs as shown in Fig.~\ref{SI_QSMWF}(b)-(d). Each
two quasi-Majorana (e.g. $\gamma_A, \gamma_B$) wave functions can be obtained by $\chi_{A}(i) =1/\sqrt{2}\left[\varphi_{+\varepsilon}(i) e^{i \theta}+\varphi_{-\varepsilon}(i) e^{-i \theta}\right]$, $\chi_{B}(i)=i/{\sqrt{2}}\left[\varphi_{\epsilon}(i)e^{i \theta}-\varphi_{-\epsilon}(i)e^{-i \theta}\right]$, where $\varphi_{\pm \varepsilon}$ is the eigenstate wave function obtained from the BDG Hamiltonian in Eq.~(\ref{eq:dBDG_H}) with near-zero eigen-energy $\pm \varepsilon$.

\begin{figure}
\centering
\includegraphics[width=0.6\columnwidth]{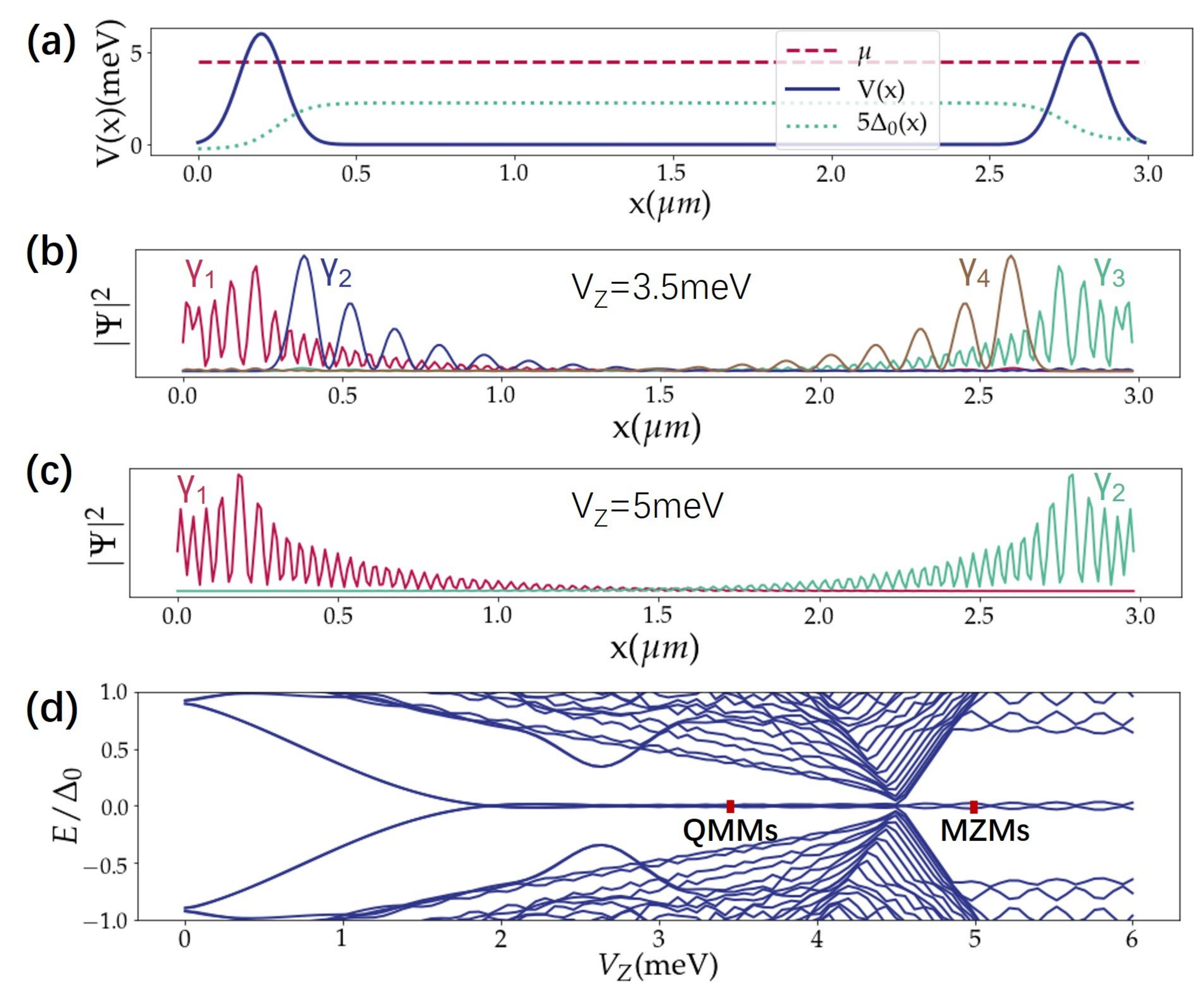}
\caption{\label{SI_QSMWF}(a) is the spatial distribution of the potential
$V\left(x\right)$ and induced gap $\Delta\left(x\right)$. (b) shows
the four quasi-Majorana wave functions along the nanowire at $V_{Z}=3.5\text{meV}$.
(c) shows the two Majorana wave functions in topological regime with
$V_{Z}=5\text{meV}$. (d) is the energy spectrum as a function of
$V_{Z}$.}
\end{figure}

\subsection{Hamiltonian of the lead electrodes}
To detect the TKE, at least three leads need to be attached to the nanowire as shown in Fig.~1(c) of the main text. The $L_1$ and $L_3$ leads are normal metallic leads which can be simulated by a one-dimensional chain, and the $L_2$ lead is ferromagnetic whose lattice Hamiltonian can be written as
\begin{equation}
\mathcal{H}_{L2}= \sum_{j}\left(\psi_{j}^{\dagger}\left[2 t_{\perp}-\mu_2-M \left(\sigma_{x} \cos\phi+\sigma_{z}\sin\phi\right) \tau_{z}\right] \psi_{j}\right.\left.-\left[\psi_{j+a_{\perp}}^{\dagger}\left(t_{\perp} \tau_{z} \right)\psi_{j}+\text { H.c. }\right]\right),\label{eq:dBDG_HLead}
\end{equation}
where $t_{\perp}=2\text{meV}$, $\mu_2=-2\text{meV}$ and $M=6meV$. $j$ labels the lattice site in the $y$ direction which means that $L_2$ lead is perpendicular to the nanowire. $j+a_{\perp}$ labels the nearest neighbour site in the $+y$ direction. $t_{\perp}$ is the hopping constant in $L_2$ lead, and a smaller hopping $t_{\perp}$ corresponds to a relatively large DOS in the lead. The angle $\phi$ represents the magnetic direction in the $x$-$z$ plane, e.g. $\phi=0$ means the the magnetization is at $+x$ direction. The $L_2$ lead is attached to the nanowire by adding a hooping term between the site $i=\text{int}\left(x_{T}\right)$ of the nanowire and the site $j=1$ of $L_2$ lead. The hopping strength $\tilde{t}$ represents the coupling strength between the wire and the lead, and $\tilde{t}$ can be controlled by tuning the tunnel barrier. Taking $\tilde{t}_0=2\text{meV}$, the quasi-Majorana wavefunctions of the cases corresponding to $\tilde{t}=0.5\tilde{t}_0,0.7\tilde{t}_0,0.9\tilde{t}_0$ are shown in Fig.~\ref{TWFchanget}. The leakage of the wavefunction of $\gamma_2$ to $L_2$ lead can be suppressed by lowering $\tilde{t}$.

\begin{figure}
\centering
\includegraphics[width=0.6\columnwidth]{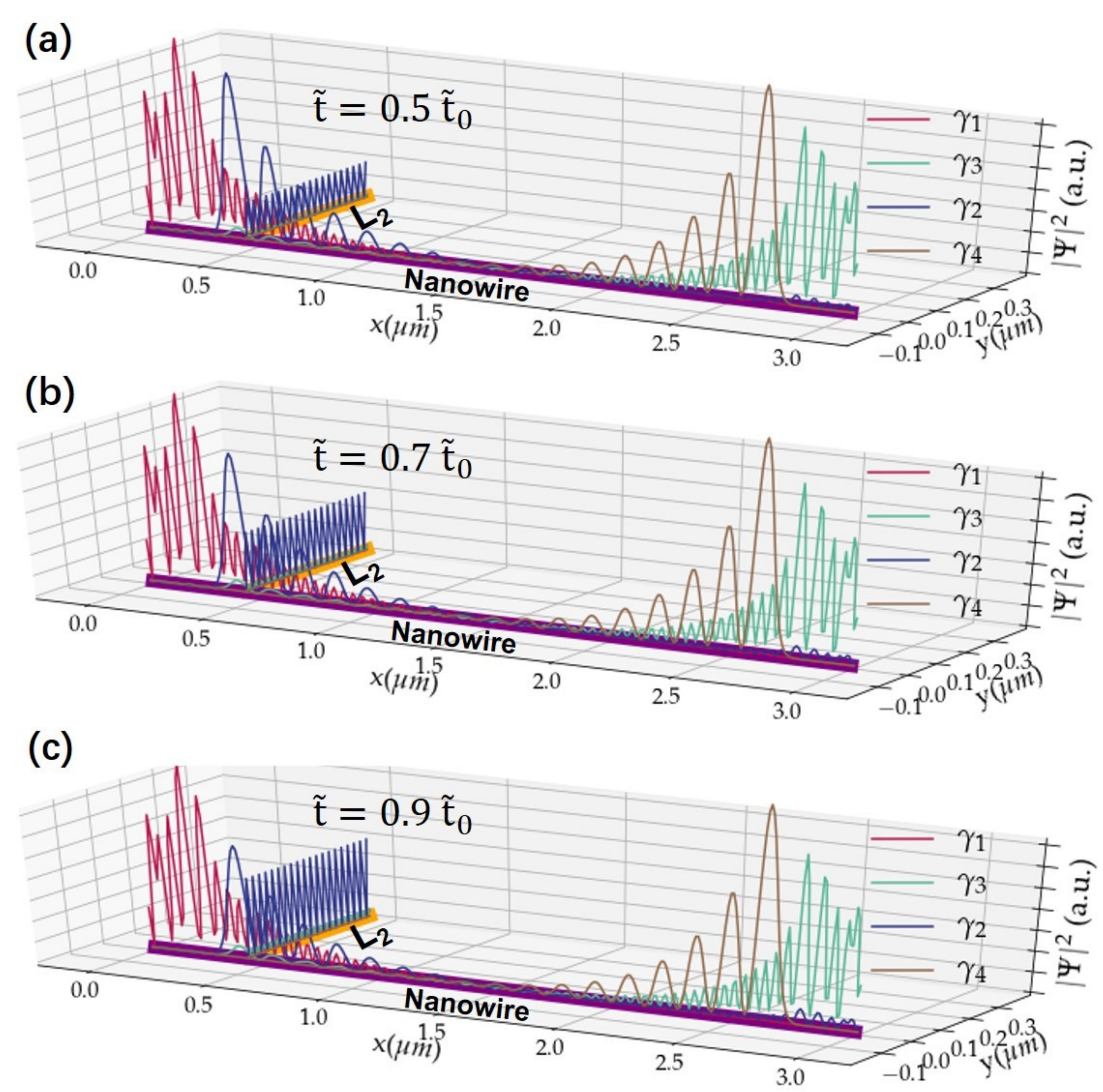}
\caption{\label{TWFchanget}The wave functions of the four QMMs $\gamma_{1-4}$ when attaching $L_2$ lead. The leakage of $\gamma_2$ to the lead is strong and can be controlled by $\tilde{t}$.}
\end{figure}

\begin{figure}
\centering
\includegraphics[width=0.7\columnwidth]{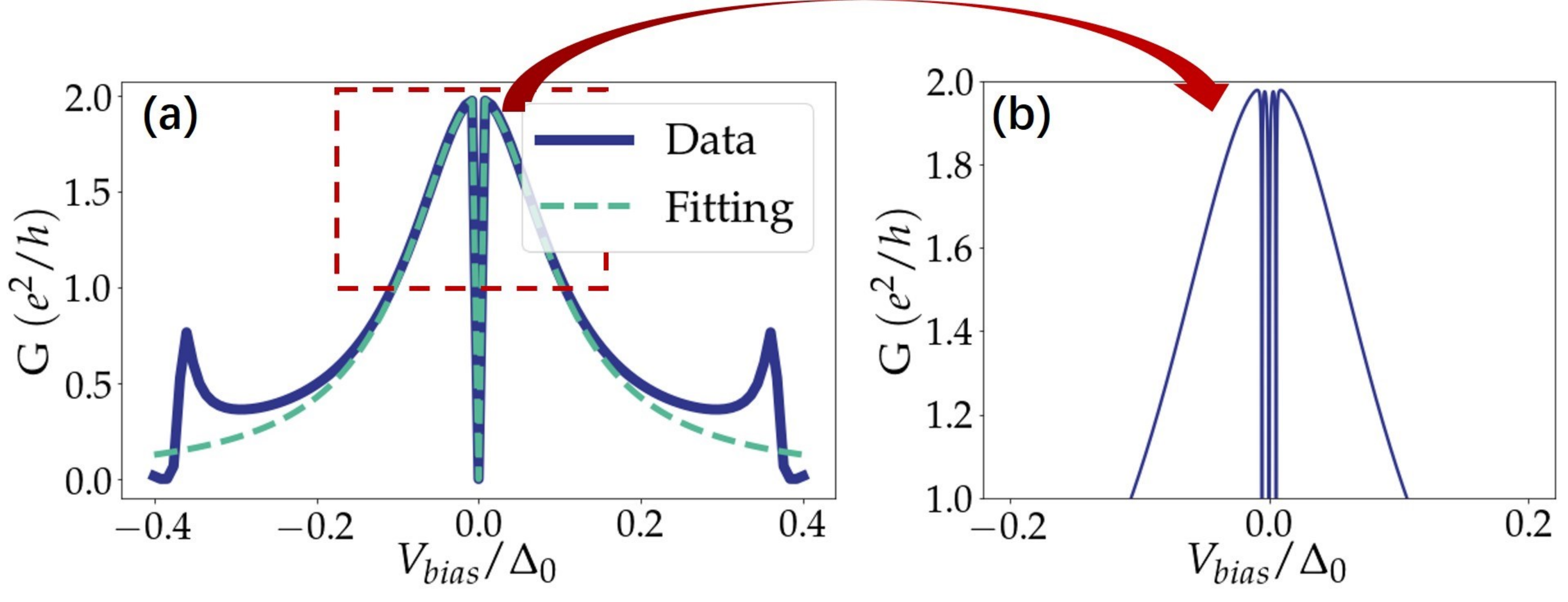}
\caption{\label{FitTunelG}(a) shows the single-terminal conductance detected by $L_2$ lead connected at $x_T = 4,2\mu m$ (solid blue) and the fitting conductance using Eq.~\ref{eq:GandGamma} as a function of the bias voltage. (b) is a zoom view of the simulated conductance peak in (a).}
\end{figure}

\section{The tunneling conductance to the ferromagnetic lead\label{section02}}
Next, we will show the details about how we evaluate the Majorana-lead and Majorana-Majorana hybridization parameters when leads are attached in our device. As mentioned in the main text, the hybridization factors $\Gamma_{12}$, $\Gamma_{22}$ and $\varepsilon_{12}$ are most influential and are simultaneously disturbed by the $L_2$ lead. To evaluate those coupling parameters, we ground the nanowire, attach the $L_2$ lead near $\gamma_2$, and numerically compute the single-terminal tunneling conductance. Those parameters can then be extracted from those numerical results. Here, we assume that those hybridization couplings are the same even if we add a finite charging energy when considering a floating nanowire island.
If the nanowire is sufficiently long, the hybridizations between the left QMMs $\gamma _{1,2}$ and the right  QMMs $\gamma _{3,4}$ can be neglected.
Under Majorana basis, the effective Hamiltonian of two Majoranas with finite hybridization $\varepsilon_{12}$ is
\begin{align}
H^M_{\mathrm{eff}}=\left(\begin{array}{cc}
0 & i \varepsilon_{12} \\
-i \varepsilon_{12} & 0
\end{array}\right).
\end{align}
With a ferromagnetic lead attached near $\gamma _2$ , the effective coupling matrix between the lead  and $\gamma _1$, $\gamma _2$ can be written as 
\begin{align}
W=\left(\begin{array}{cc}
\tau_{12} & -\tau_{12}^{*} \\
\tau_{22} & -\tau_{22}^{*}
\end{array}\right),
\end{align}
where the $\tau_{12}$, $\tau_{22}$ are the effective couplings between $L_2$ lead and $\gamma_{1,2}$, and the two columns of the matrix $W$ represent the electron part and the hole part of the $L_2$ lead respectively. With the Mahaux-Weidenm\"uller formula we can obtain the scattering matrix
of the junction between nanowire and $L_2$ lead:

\begin{align}
S(\omega)=1-2 \pi i W^{\dagger}\left(\omega-H^M_{\mathrm{eff}}+i \pi W W^{\dagger}\right)^{-1} W.\label{eq:SandW}
\end{align}
Then the single-terminal tunneling conductance in the $L_2$ lead when applying a bias voltage $V$ ($V<\Delta _0$) is $G(V)=\frac{2 e^{2}}{h} \operatorname{Tr} S_{h e}^{\dagger}(eV) S_{h e}(eV)$. Using Eq. (\ref{eq:SandW}), we can obtain the conductance:
\begin{align}
G(V)=\frac{2 e^{2}}{h} \frac{\left(\Gamma_{12}-\Gamma_{22}\right)^{2} (eV)^{2}}{\left[(eV)^{2}-\varepsilon_{12}^{2}-\Gamma_{12} \Gamma_{22}\right]^{2}+\left(\Gamma_{12}+\Gamma_{22}\right)^{2} (eV)^{2}},\label{eq:GandGamma}
\end{align}
where $\Gamma_{ij}=2\pi\left|\tau_{ij}\right|^{2}$, and $\tau_{21}$ and $\tau_{22}$ has a phase difference $\pi/2$. A curve fitted by Eq. (\ref{eq:GandGamma}) is shown in Fig.~\ref{FitTunelG}(a). The blue curve in Fig.~\ref{FitTunelG}(a) is obtained from the Kwant simulation using the set-up given by Sec.~\ref{section02} with $\tilde{t}=\tilde{t}_0$. The zoom view of the conductance peak shows that there is also a narrow split peak, which is contributed by the two QMMs on the other side of the nanowire. Nevertheless, their couplings to this lead are too weak, and we can neglect their influence. In the main text, we assume that the nanowire is long enough, and we only consider the Majorana-Majorana and Majorana-lead hybridizations at the same end.

By changing the location $x$ of the connection point of the $L_2$ lead,  we can obtain a series of hybridization parameters with Eq.~(\ref{eq:GandGamma}) as shown in Fig.~3(d) of the main text. The spatial variations of $\Gamma_{22}$ and $\Gamma_{12}$ are related to the spatial distributions of the wave functions and spin densities of $\gamma_{2}$ and $\gamma_{1}$ as shown in Fig.~\ref{WFunctions}. Given the spinor representation of the particle-hole symmetric quasi Majorana wave function $\chi_\alpha (i)=\left({u}_{\alpha i \uparrow}, {u}_{\alpha i \downarrow}, {u}_{\alpha i \uparrow}^{*}, {u}_{\alpha i \downarrow}^{*}\right)^{T}$, the spin density of $\gamma_{\alpha}$ for different directions is obtained from the formula $\left\langle \sigma_{\nu}\right\rangle_{\alpha}(i)= \sum_{s, s^{\prime}} {u}_{\alpha i s}^{*}\left[\sigma_{\nu}\right]_{s s^{\prime}} {u}_{\alpha i s^{\prime}}$. From Fig.~\ref{WFunctions} we can see that the two QMMs have opposite spin directions when projecting onto the $\sigma_x$ eigen-basis. The two QMMs also have nonzero $\left\langle \sigma_{z}\right\rangle_\alpha$ which is much smaller than $\left\langle \sigma_{x}\right\rangle_\alpha$. The small $\left\langle \sigma_{z}\right\rangle_\alpha$ will result in a small asymmetry of the value $(T_K-T_h)/T_h$ around $\varphi=0$ shown in Fig.~3(b) of the main text. 
\begin{figure}
\centering
\includegraphics[width=0.5\columnwidth]{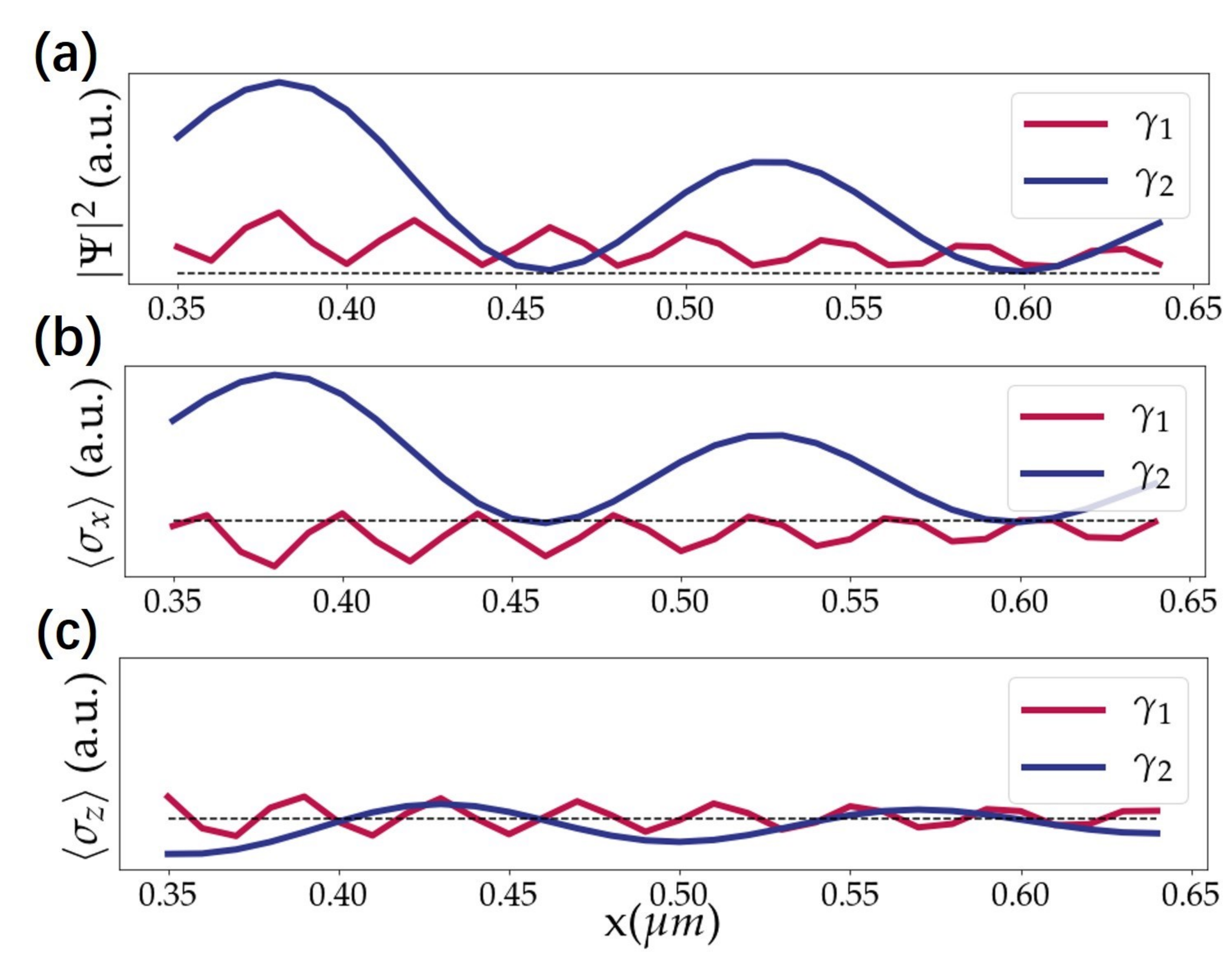}
\caption{\label{WFunctions}The wave functions and the spin densities of the two QMMs $\gamma_1$ and $\gamma_2$. The ranges of $y$-axis of (a) and (b) are the same which shows that the spin is mainly in the $\pm x$ direction.}
\end{figure}
With a fixed connection point $x_T$ between $L_2$ lead and the nanowire, the wave functions of QMMs can be shifted horizontally by changing the chemical potential $\mu$ or the Zeeman field $V_Z$; therefore, the hybridization can be tuned by changing the chemical potential or the Zeeman field, as well. A comparison of the hybridization couplings $\Gamma_{22}$, $\Gamma_{12}$ and $\varepsilon_{12}$ for changing the control parameters $x_T$, $\mu$, $V_Z$ is shown in Fig. \ref{Gammas}. Therefore, in a practical situation with a fixed $x_T$, the values of hybridization parameters can be tuned by changing the chemical potential or the magnetic field.
\begin{figure}
\centering
\includegraphics[width=0.9\columnwidth]{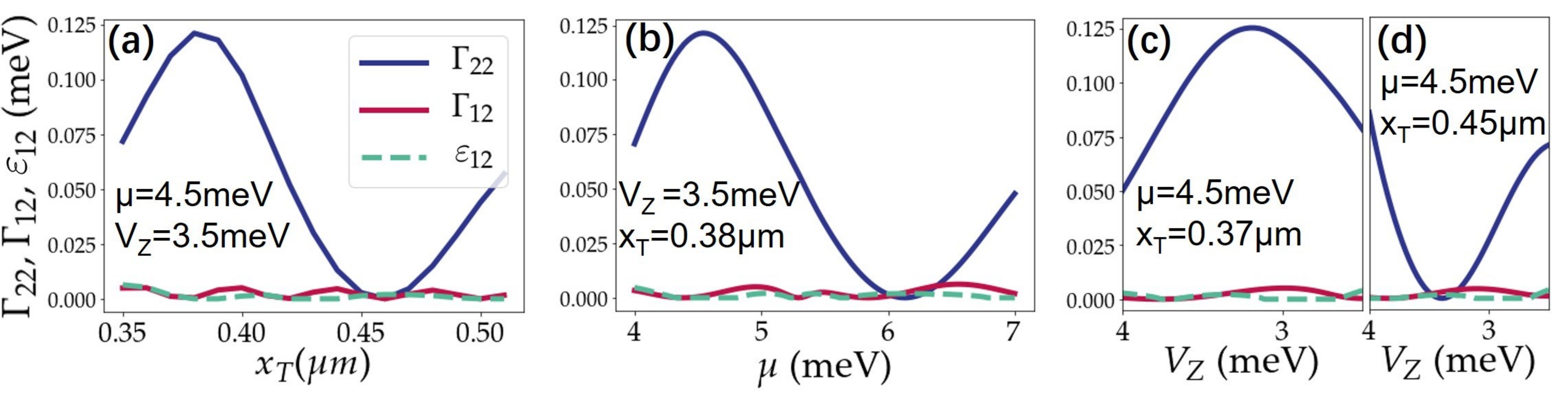}
\caption{\label{Gammas} The variations of hybridization parameters $\Gamma_{22}$, $\Gamma_{12}$ and $\varepsilon_{12}$ as a function of $x_T$, $\mu$ and $V_Z$.}
\end{figure}

\end{widetext}

\bibliographystyle{apsrev4-1} 
\bibliography{TKEinQM}

\end{document}